\documentclass[a4paper,12pt]{article}
\usepackage[margin=1.3cm]{geometry}
\usepackage{comment}
\usepackage{amssymb,extarrows,graphicx,subfigure,setspace}
\usepackage{cite}
\usepackage{slashed}
\usepackage{tensor}
\usepackage[toc,page]{appendix}
\usepackage{color}
\usepackage{physics}
\usepackage{hyperref}
\usepackage{dirtytalk}
\hypersetup{colorlinks=true, linkcolor=blue, citecolor=red, linktoc=page}
\makeatother

\usepackage{float}
\usepackage{textcomp}
\usepackage{amsmath}
\newmuskip\pFqmuskip

\newcommand*\pFq[6][8]{%
  \begingroup 
  \pFqmuskip=#1mu\relax
  \mathcode`\,=\string"8000
  \begingroup\lccode`\~=`\,
  \lowercase{\endgroup\let~}\pFqcomma
  {}_{#2}F_{#3}{\left[\genfrac..{0pt}{}{#4}{#5};#6\right]}%
  \endgroup
}
\newcommand{\pFqcomma}{\mskip\pFqmuskip}

\usepackage{mathrsfs}
\usepackage{hyperref}

\newcommand{\be}{\begin{equation}}
\newcommand{\bea}{\begin{eqnarray}}
\newcommand{\eea}{\end{eqnarray}}
\newcommand{\ba}{\begin{array}}
\newcommand{\ea}{\end{array}}
\newcommand{\ee}{\end{equation}}
\newcommand{\bes}{\begin{equation*}}
\newcommand{\beas}{\begin{eqnarray*}}
\newcommand{\eeas}{\end{eqnarray*}}
\newcommand{\bas}{\begin{array*}}
\newcommand{\eas}{\end{array*}}
\newcommand{\ees}{\end{equation*}}

\numberwithin{equation}{section}

\begin{document}
\color{black}
\begin{center}
\Large{\bf Weak Gravity Conjecture Validation with Photon Spheres of Quantum Corrected AdS-Reissner-Nordstrom Black Holes in Kiselev Spacetime }\\
\small \vspace{1cm}
 {\bf Mohammad Reza Alipour $^{a,b}$\footnote {Email:~~~mr.alipour@stu.umz.ac.ir}},\quad
 {\bf Mohammad Ali S. Afshar $^{b}$\footnote {Email:~~~m.a.s.afshar@gmail.com}},\quad\\\vspace{0.25cm}
 {\bf Saeed Noori Gashti$^{b}$\footnote {Email:~~~saeed.noorigashti@stu.umz.ac.ir}}, \quad
 {\bf Jafar Sadeghi$^{a}$\footnote {Email:~~~pouriya@ipm.ir}} \quad\\
\vspace{0.5cm}$^{a}${Department of Physics, Faculty of Basic
Sciences,
University of Mazandaran\\
P. O. Box 47416-95447, Babolsar, Iran}\\
\vspace{0.5cm}$^{b}${School of Physics, Damghan University, P. O. Box 3671641167, Damghan, Iran}\\
\small \vspace{1cm}
\end{center}
\begin{abstract}
In this study, we investigate the Weak Gravity Conjecture (WGC) in the context of quantum-corrected AdS-Reissner-Nordström (AdS-RN) black holes within Kiselev spacetime. Our focus is on photon spheres, which serve as markers for stable and unstable photon spheres. We confirm the validity of the WGC by demonstrating that quantum corrections do not alter the essential charge-to-mass ratio, thereby supporting the conjecture's universality. Our analysis reveals that black holes with a charge greater than their mass ($Q > M$) possess photon spheres or exhibit a total topological charge of the photon sphere (PS = -1), which upholds the WGC. This finding is significant as it reinforces the conjecture's applicability even in the presence of quantum corrections. Furthermore, we examine various parameter configurations to understand their impact on the WGC. Specifically, we find that configurations with $\omega = -\frac{1}{3}$ and $\omega = -1$ maintain the conjecture, indicating that these values do not disrupt the charge-to-mass ratio required by the WGC. However, for $\omega = -\frac{4}{3}$, the conjecture does not hold, suggesting that this particular parameter value leads to deviations from the expected behavior. These results open new directions for research in quantum gravity, as they highlight the importance of specific parameter values in maintaining the WGC. The findings suggest that while the WGC is robust under certain conditions, there are scenarios where it may be challenged, prompting further investigation into the underlying principles of quantum gravity.
\end{abstract}
\tableofcontents
\newpage
\section{Introduction}
The use of quantum corrections addresses the singularity problem in classical general relativity. In the classical Schwarzschild solution, there is a singularity at \( r = 0 \), where the curvature of spacetime becomes infinite. D.I. Kazakov and S.N. Solodukhin\cite{1} consider spherically symmetric quantum fluctuations of the metric and the matter fields, which lead to an effective two-dimensional dilaton gravity model. The quantum corrections deform the Schwarzschild solution in such a way that the classical singularity at \( r = 0 \) is replaced by a quantum-corrected region. This region has a minimum radius \( r_{\text{min}} \) of the order of the Planck length \( r_{\text{Pl}} \), and the scalar curvature remains finite. The results of this correction are significant because they suggest a spacetime that is regular and free of singularities\cite{1}. The space-time consists of two asymptotically flat sheets that are glued at a hypersurface of constant radius. This implies that quantum effects can potentially smooth out the singularities predicted by classical general relativity, leading to a more complete understanding of black hole physics and spacetime structure at the quantum level. Therefore, the exploration of quantum corrections has garnered attention from other researchers, leading to studies in areas such as the criticality and efficiency of black holes\cite{2}, the thermodynamics of a quantum-corrected Schwarzschild black hole surrounded by quintessence\cite{3}, accretion onto a Schwarzschild black hole against a quintessence background\cite{4}, as well as investigations into quasinormal modes, scattering, shadows\cite{5}, and the Joule-Thomson expansion\cite{6}
\\\\
The (WGC) is a theoretical framework in the field of quantum gravity. It posits that in any consistent theory of quantum gravity, there must exist particles for which the gauge forces are stronger than gravitational forces. This conjecture has profound implications across various domains of physics and mathematics. A key aspect of the WGC is its potential to connect different areas of theoretical physics. It suggests that gravity should be the weakest force, which has led to several variations of the conjecture. These variations have been found to be consistent with all known quantum gravity theories, particularly within the string theory landscape. For a more comprehensive understanding, you might find\cite{a,b,f,h,i,j,k,l,n,o,p,t,u,w,x,ee,hh,ii,mm,ss,xx,zz,aaa,bbb,ccc,ddd,fff,hhh,jjj,kkk,lll} that provides an extensive discussion on the various swampland conjecture, their variations, and their implications for particle physics, cosmology, general relativity, and mathematics.\\\\
The presence of a photon sphere is a critical aspect of ultra-compact gravitational entities such as black holes, and its necessity is well-supported by extensive research across a variety of black hole models\cite{7,8,9,10,11,12,13,14,15,16,17,18,19,20,21}. This fundamental feature opens the door to numerous intriguing consequences. For example, unstable spherical photon spheres will dominate space outside the event horizon when the weak cosmic censoring conjecture (WCCC) holds and the structure is in the form of a black hole, so probing the gravitational structure under WGC conditions to find unstable spheres can be a special advantage. This is because an unstable photon sphere predominates the spatial domain of the system only in the presence of an event horizon, thereby fulfilling the WCCC criterion. Moreover, the most profound implication, which constitutes the central objective of our study, revolves around the ability to categorize the parameter space of the examined models. This categorization hinges on the presence and specific locations of both topological photon spheres and their counterparts, anti-photon spheres. Consequently, we can delineate with precision the parameter ranges that align with a black hole configuration, a naked singularity\cite{22,22'}.\\

We are going to pursue a goal that is a novel and new process in examining the approach to the relation of the cosmos and quantum. In our study, we delve into the studieng of the (WGC) against the backdrop of quantum-enhanced (AdS-RN) black holes situated within Kiselev spacetime, placing a spotlight on the pivotal role of photon spheres. Through the integration of quantum corrections and the unique framework of Kiselev spacetime, we pinpoint the extremality conditions of these black holes. Our research corroborates the WGC, broadening its scope to encompass the intricate black hole configurations found in Kiselev spacetime. We ascertain that, notwithstanding the quantum adjustments, the charge-to-mass ratio consistently exceeds the WGC's anticipated threshold, thus affirming the conjecture's widespread applicability. Additionally, our investigation illuminates the complex synergy between quantum dynamics and gravitational forces, paving the way for novel explorations in the realms of high-energy physics and quantum gravity. \\
Building upon the concepts discussed above, we have structured the article as follows: Section 2 introduces the model under consideration viz the quantum-corrected AdS-RN black holes within Kiselev spacetime. In Section 3, we explore the photon sphere structure of the aforementioned black hole to derive the weak gravity conjecture and the weak cosmic censorship conjecture, considering the model's free parameters. We will examine these results in detail. Finally, Section 4 presents the conclusions of our study.
\section{The Model}
Here, we will meticulously detail the quantum-corrected AdS-RN black hole which is enveloped by Kiselev spacetime. Our focus will be to conduct a comprehensive analysis of its characteristics with photon spheres to prove the WGC that holds significance for our research. We will delve into the spacetime metric of this quantum-corrected charged AdS black hole, which is surrounded by a cosmological fluid. This metric is characterized by its spherical symmetry and is expressed as follows\cite{6},
\begin{equation}\label{Ph1}
\begin{split}
ds^2=f(r)dt^2-f(r)^{-1}dr^2-r^2d\Omega^2,
\end{split}
\end{equation}
So, one can obtain\cite{6},
\begin{equation}\label{Ph2}
\begin{split}
f(r)=-\frac{2M}{r}+\frac{\sqrt{r^2-a^2}}{r}+\frac{r^2}{\ell^2}-\frac{c}{r^{3\omega+1}}+\frac{Q^2}{r^2},
\end{split}
\end{equation}
In our study, we will elucidate the parameters defining the black hole under investigation. The parameter \( M \) represents the mass of the black hole, while \( a \) is associated with the quantum corrections applied to the black hole's properties. The symbol \( \ell \) denotes the length scale pertinent to the asymptotically AdS (Anti-de Sitter) spacetime. The parameter \( c \) correlates with the cosmological fluid surrounding the black hole, and \( Q \) signifies the electric charge of the black hole. To commence our discussion, it is essential to understand the rationale behind our selection of the aforementioned metric and to dissect the genesis of each term within it. Notably, M. Visser has recently put forward the notion that the Kiselev black hole model can be extended to a spacetime comprising \( N \) components. This generalization is characterized by a linear relationship between energy and pressure for each individual component, as delineated in the literature\cite{23,24}. For more study about quantum corrected charged AdS black hole surrounded by Kiselev spacetime, you can see\cite{6}. In this study, we consider some values for $\omega $ such as $ \omega = -\frac{1}{3} $, $ -\frac{2}{3} $, -1 and $ \omega = -\frac{4}{3} $. The parameter \( a \) is intricately linked to modifications in the black hole's mass, which arise due to quantum corrections. The foundational theory behind this parameter is thoroughly discussed in reference\cite{25}. As an independent variable, \( a \) holds the unique property that when set to zero, the metric simplifies to the well-known AdS-Reissner-Nordström metric, now enveloped by a cosmic fluid. Theoretically, \( a \) can adopt any value, provided it remains less than the event horizon's radius. This stipulation aligns with the expectation that \( a \) represents a minor adjustment to the established black hole metric. The first law of thermodynamics is sufficiently robust to include variations in the defining parameters, or 'hair,' of the black hole. In this context, the parameters in question are the black hole's area, the cosmological constant, the electric charge, the quintessence parameter, and the quantum correction parameter. For an in-depth discussion on the incorporation of the quintessence parameter as a thermodynamic variable, please refer to \cite{26}. Building upon this, our study extends the framework to accommodate the variability of the quantum correction parameter.
\begin{equation}\label{Ph3}
\begin{split}
dM=TdS+VdP+\phi dQ+\mathcal{C}dc+\mathcal{A}da,
\end{split}
\end{equation}
To calculate the entropy of the quantum-corrected Schwarzschild black hole situated within Kiselev spacetime, reference\cite{4} reveals that this entropy aligns with the Hawking-Bekenstein entropy formulation. Consequently,
\begin{equation}\label{Ph4}
\begin{split}
&S=\frac{A}{4}=\frac{4\pi r_+^2}{4}\\
&r_+=\sqrt{\frac{S}{\pi}},
\end{split}
\end{equation}
Concerning the final thermodynamic variable, pressure, we will treat it as being associated with the cosmological constant. Consequently, the expression for pressure is $P=\frac{3}{8\pi\ell^2}$\cite{27}. The mass and Hawking temperature of the quantum-corrected AdS-RN black hole which is enveloped by Kiselev spacetime are determined as follows,
\begin{equation}\label{Ph5}
\begin{split}
M=\frac{1}{2\sqrt{\pi}}\bigg(\sqrt{S-\pi a^2}-c\pi^{\frac{3\omega+1}{2}}S^{-\frac{3\omega}{2}}+\frac{8PS^{3/2}}{3}+\pi Q^2S^{-1/2}\bigg),
\end{split}
\end{equation}
and
\begin{equation}\label{Ph6}
\begin{split}
T_{H}=\big(\frac{\partial M}{\partial S}\big)_{P,Q}=\frac{1}{4\sqrt{\pi}}\bigg(\frac{1}{\sqrt{S-\pi a^2}}+8P\sqrt{S}+3\frac{c\omega}{\sqrt{\pi}}\big(\frac{\pi}{S}\big)^{\frac{3\omega}{2}+1}-\frac{\pi Q^2}{S^{3/2}}\bigg),
\end{split}
\end{equation}

\begin{figure}[h!]
 \begin{center}
 \subfigure[]{
 \includegraphics[height=5.5cm,width=6.5cm]{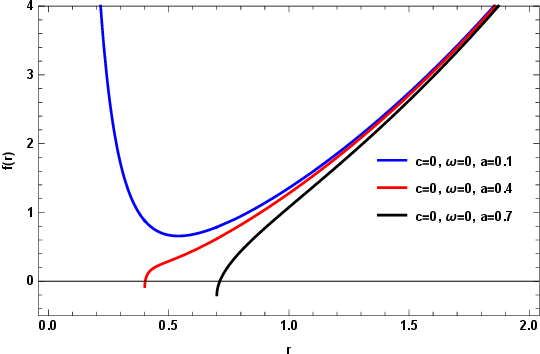}
 \label{1a}}
 \subfigure[]{
 \includegraphics[height=5.5cm,width=6.5cm]{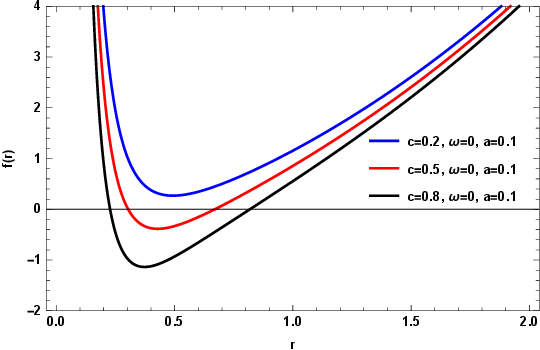}
 \label{1b}}
 \subfigure[]{
 \includegraphics[height=5.5cm,width=6.5cm]{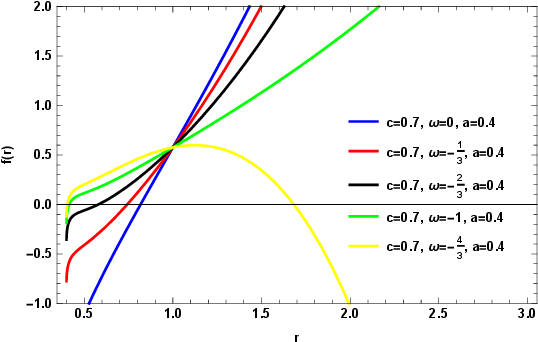}
 \label{1c}}
\subfigure[]{
 \includegraphics[height=5.5cm,width=6.5cm]{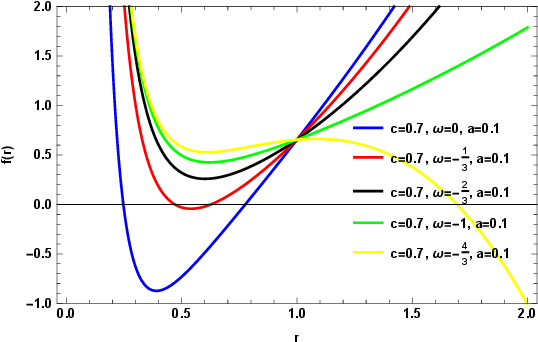}
 \label{1d}}
  \caption{\small{The plot of $f(r)-r$ for quantum-corrected (AdS-RN) black holes in Kiselev spacetime with respect to free parameters }}
 \label{1}
 \end{center}
 \end{figure}
In Fig. (\ref{1}), the changes in the metric function of the quantum-corrected (AdS-RN) black holes in the Kiselev spacetime  are plotted for different free parameters.
\newpage
\section{Methodology and Discussion}
We want to examine the (WGC) in the setting of quantum-corrected AdS-Reissner-Nordström black holes within Kiselev spacetime, focusing on photon spheres that affect the black holes' charge-to-mass ratio. Our analysis of the extremality conditions suggests that these black holes can evaporate without leaving an extremal remnant or a naked singularity. So with respect to\cite{21,22,22',23}, we start with a regular potential,
\begin{equation}\label{Ph7}
\begin{split}
H(r,\theta)=\sqrt{\frac{-g_{tt}}{g_{\varphi\varphi}}}=\frac{1}{\sin\theta}\bigg(\frac{f(r)}{h(r)}\bigg)^{1/2},
\end{split}
\end{equation}
The examination of the potential will enable us to determine the radius of the photon sphere, which is located at,
\begin{equation}\label{Ph7}
\begin{split}
\partial_r H=0,
\end{split}
\end{equation}
Therefore, we can employ a vector field denoted by $ \phi = (\phi^r, \phi^\theta) $. So, we will have,
\begin{equation}\label{Ph8}
\begin{split}
&\phi^r=\frac{\partial_r H}{\sqrt{g_{rr}}}=\sqrt{g(r)}\partial_r H,\\
&\phi^\theta=\frac{\partial_\theta H}{\sqrt{g_{\theta\theta}}}=\frac{\partial_\theta H}{\sqrt{h(r)}},
\end{split}
\end{equation}
Additionally, the winding number can be determined using the formula below,
\begin{equation}\label{Ph9}
\begin{split}
\omega_i=\frac{1}{2\pi}\int_{C_i}d\Lambda,
\end{split}
\end{equation}
where $\Lambda=\frac{\phi^2}{\phi^1}$. So the total charge is given by,
\begin{equation}\label{Ph10}
\begin{split}
Q=\sum_{i}\omega_i,
\end{split}
\end{equation}
In conclusion, the presence of a zero point within a closed curve indicates that the charge \( Q \) is precisely equivalent to the winding number. Given that every black hole likely possesses photon spheres, and considering that topological flow manifests solely at the zero points of the vector field—which delineates the photon sphere's location—it is feasible to attribute a distinct topological charge to each photon sphere. This charge may assume a value of either -1 or +1\cite{21}. Furthermore, depending on the selection of the total closed curve, which may encompass none, one, or several zero points, the total charges can also be -1, 0, or +1, in accordance with\cite{21}. Here, we will provide a detailed explanation of our computational approach for analyzing the special black hole. By applying the theoretical concepts discussed earlier and utilizing Equations (\ref{Ph2}), (\ref{Ph7}), and (\ref{Ph8}), we derive the following results.
\begin{equation}\label{Ph11}
\begin{split}
\phi^r=\bigg[\csc (\theta ) \left(r \left(\frac{3 a^2-2 r^2}{\sqrt{r^2-a^2}}+3 c (w+1) r^{-3 w}+6 M\right)-4 Q^2\right)\bigg]\times\bigg[2 r^4\bigg]^{-1}
\end{split}
\end{equation}
and
\begin{equation}\label{Ph12}
\begin{split}
\phi^\theta=\bigg[\cot (\theta ) \csc (\theta ) \sqrt{\frac{r \left(\sqrt{r^2-a^2}-c r^{-3 w}+\frac{r^3}{l^2}-2 M\right)+Q^2}{r^2}}\bigg]\times\bigg[r^2\bigg]^{-1}
\end{split}
\end{equation}
We seek new evidence to substantiate WGC. The most significant innovation in this article is the identification of regions where, while these conjectures hold, the survival of black hole structures with the presence of unstable photon spheres is confirmed. This is a distinctive feature of black hole structures.
An even more intriguing point to consider is that the WGC arises from quantum structures, whereas photon spheres are generated by relativistic structures. Bridging these concepts serves as a starting point for connecting quantum mechanics and relativity, providing a dual motivation for this work.
The WGC suggests that in any consistent theory of quantum gravity, gravity is the weakest force. It implies that there must be a particle with a charge-to-mass ratio greater than that of an extremal black hole. An extremal black hole has the maximum possible charge or angular momentum for a given mass, such that its temperature is zero. The WGC is often tested using extremal black holes because they provide a natural setting to explore the balance between gravitational and other forces. Based on these concepts, the WGC suggests that under certain conditions, which could be due to environmental factors, a situation might arise in limited regions around a black hole where the charge-to-mass ratio exceeds one.\\
The question is, in such a situation, can the studied structure still maintain its black hole shape?
To address this question, we propose an innovative approach by examining the photon sphere for such a structure using topological methods. Ultra-compact gravitational structures, of which black holes are a valuable and study-worthy example, cause photon deflection, creating conditions for the formation of photon rings and spheres. Additionally, studies have shown that by rewriting the effective potential and using mapping to a two-dimensional plane, a topological charge can be assigned to each photon sphere. Multiple studies on black holes with various structures have confirmed that black holes in their natural form, while simultaneously adhering to the WCCC (having an event horizon), must also possess an unstable photon sphere, which is the basis for studying the black hole shadow. Numerous studies on photon spheres based on topological charge have shown that the charge of such a structure will be -1.\\\\

\begin{figure}[h!]
 \begin{center}
 \subfigure[]{
 \includegraphics[height=4.5cm,width=5cm]{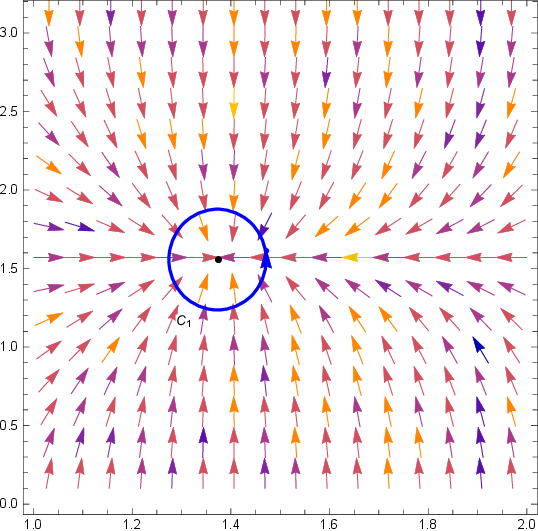}
 \label{2a}}
 \subfigure[]{
 \includegraphics[height=4.5cm,width=5cm]{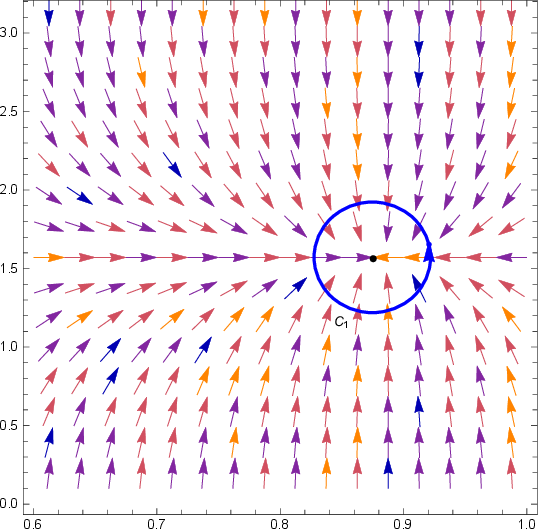}
 \label{2b}}
 \subfigure[]{
 \includegraphics[height=4.5cm,width=5cm]{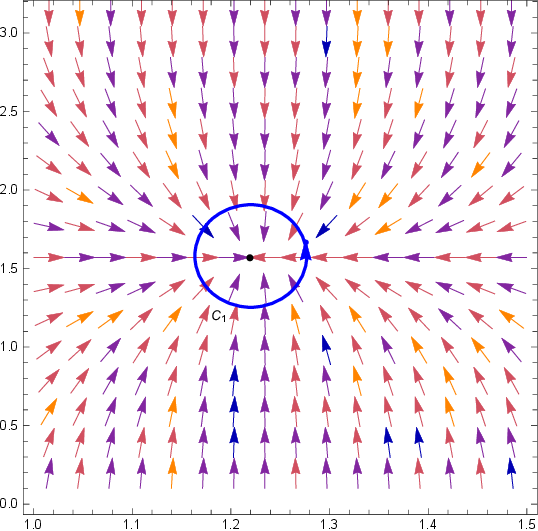}
 \label{2c}}
  \caption{\small{The plot of photon spheres with respect to $ c = 0, \ell = 1,  a = 0.4$ for $M>Q$ in Fig. (\ref{2a}), M=Q in Fig. (\ref{2b}) and $Q>M$ in Fig. (\ref{2c})}}
 \label{2}
 \end{center}
 \end{figure}
The WGC suggests that extremal black holes should be unstable to decay into smaller black holes or other particles with higher charge-to-mass ratios. This instability helps ensure that gravity remains the weakest force. In summary, we convert the black hole into the extremal state with WGC, which plays a crucial role in testing the Weak Gravity Conjecture and exploring the connections between quantum mechanics and general relativity. Regarding the first part of the review, we chose certain points that have a very clear rationale. As we aim to confirm the Weak Gravity Conjecture and find supporting evidence, identifying even a single compatible point through various cosmological concepts is valuable. Points consistent with the Weak Gravity Conjecture can serve as a starting point for further investigations. By locating these points of compatibility, we seek to determine the intervals in which quantum gravity is compatible. In coupled theories of gravity and quantum gravity, we understand that the form cannot be global but must be local. We examine whether this compatibility with the Weak Gravity Conjecture is maintained and in what intervals this consistency exists, as briefly stated in the paper's table. Additionally, by proving this theorem, the structure of photon spheres can be used to investigate other aspects of black hole structures, providing further evidence for the WGC.\\

\begin{figure}[h!]
 \begin{center}
 \subfigure[]{
 \includegraphics[height=4.5cm,width=5cm]{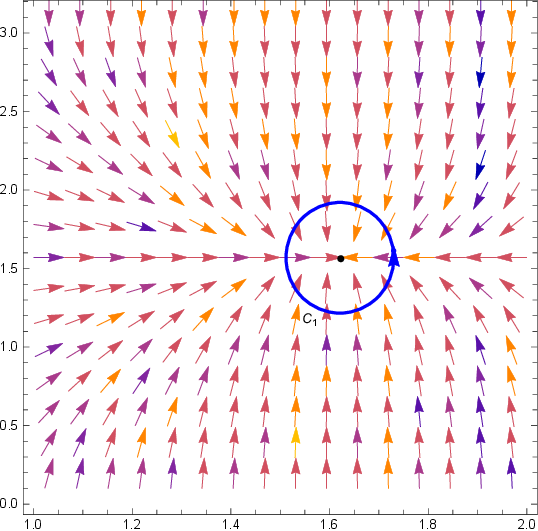}
 \label{3a}}
 \subfigure[]{
 \includegraphics[height=4.5cm,width=5cm]{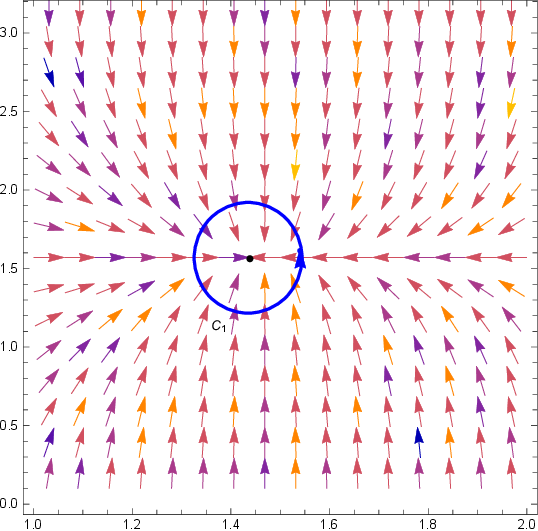}
 \label{3b}}\\
 \subfigure[]{
 \includegraphics[height=4.5cm,width=5cm]{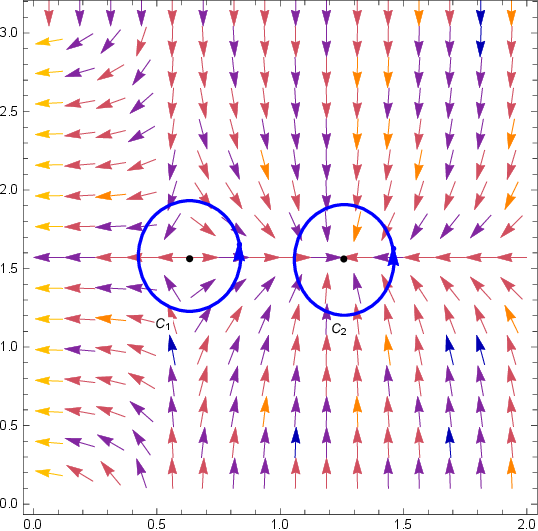}
 \label{3c}}
 \subfigure[]{
 \includegraphics[height=4.5cm,width=5cm]{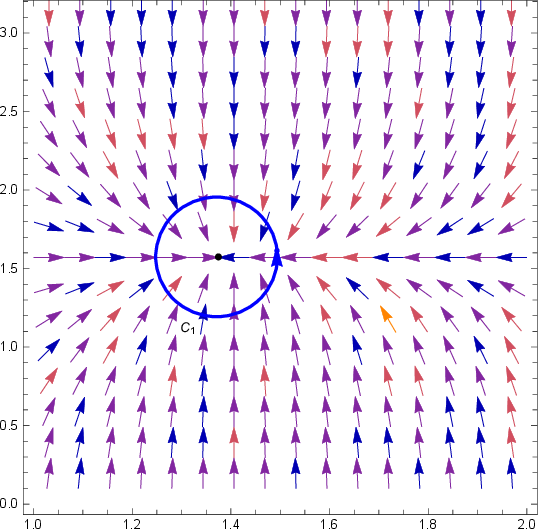}
 \label{3d}}
  \caption{\small{The plot of photon spheres with respect to $\ell = 1,  a = 0.1,  c = 0.2, \omega = 0$ for $M>Q$ in Fig. (\ref{3a}), M=Q in Fig. (\ref{3b}) and $Q>M$ in Fig. (\ref{3c}) and with respect to $\ell = 1,  a = 0.4,  c = 0.2, \omega = 0$ for $Q>M$ in Fig. (\ref{3d})}}
 \label{2}
 \end{center}
 \end{figure}

To investigate the compatibility of the WGC and the WCCC in the context of the mentioned black hole, we outline several steps. First, we should determine the conditions under which the mentioned black hole has event horizons. For a quantum-corrected AdS-Reissner-Nordström black hole with charge $Q $ and mass $ M $ in Kiselev spacetime, the conditions for the existence of event horizons are as;  Without quantum corrections and Kiselev spacetime, the black hole has two event horizons if $Q^2/M^2 \leq 1$ and none event horizon if $Q^2/M^2 > 1$, leading to a naked singularity, which violates the WCCC.
With quantum corrections and Kiselev spacetime, the event horizons depend on $Q $, $ M $, $\ell$, $c$, $\omega$, and $ a$. The singularity can be covered, satisfying the WCCC. We investigate the parameter space where both the WGC and WCCC are satisfied. This involves finding constraints on $Q $, $ M $, $\ell$, $c$, $\omega$, and $ a$ such that the black hole has event horizons WCCC and also the charge-to-mass ratio meets with the WGC criteria.
Such a situation is easily identified by examining the metric of the black hole.\\\\

 \begin{figure}[h!]
 \begin{center}
 \subfigure[]{
 \includegraphics[height=4.5cm,width=5cm]{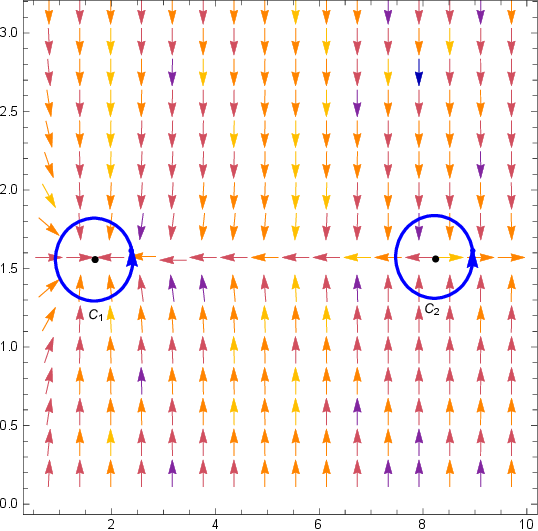}
 \label{4a}}
 \subfigure[]{
 \includegraphics[height=4.5cm,width=5cm]{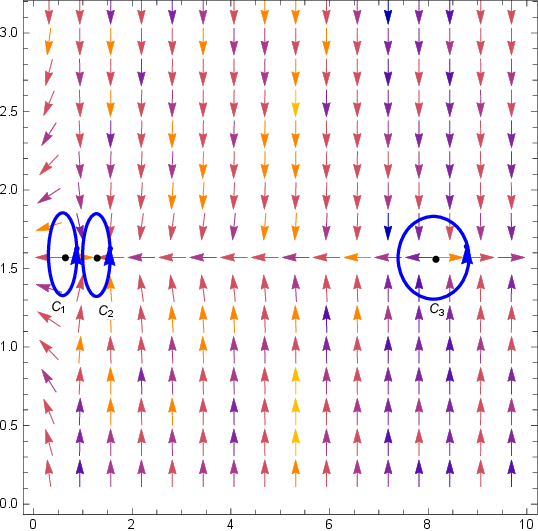}
 \label{4b}}
 \subfigure[]{
 \includegraphics[height=4.5cm,width=5cm]{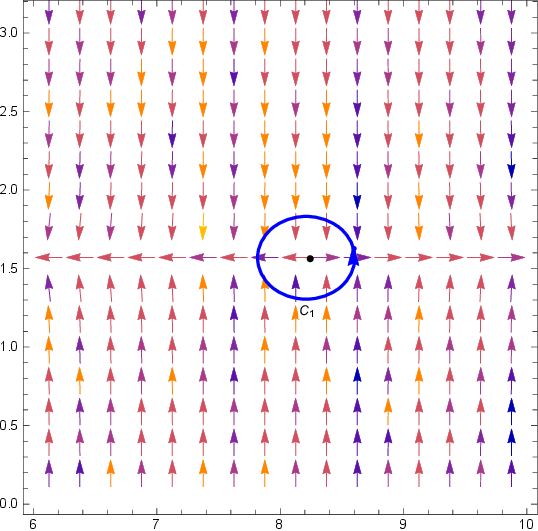}
 \label{4c}}
  \caption{\small{The plot of photon spheres with respect to $\ell = 1,  a = 0.1,  c = 0.2, \omega = -2/3$ for $M>Q$ in Fig. (\ref{4a}), M=Q in Fig. (\ref{4b}) and $Q>M$ in Fig. (\ref{4c})}}
 \label{4}
 \end{center}
 \end{figure}
\begin{figure}[h!]
 \begin{center}
 \subfigure[]{
 \includegraphics[height=4.5cm,width=5cm]{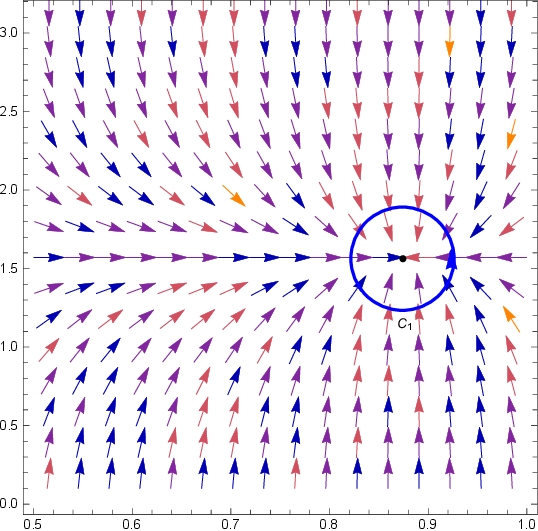}
 \label{6a}}
 \subfigure[]{
 \includegraphics[height=4.5cm,width=5cm]{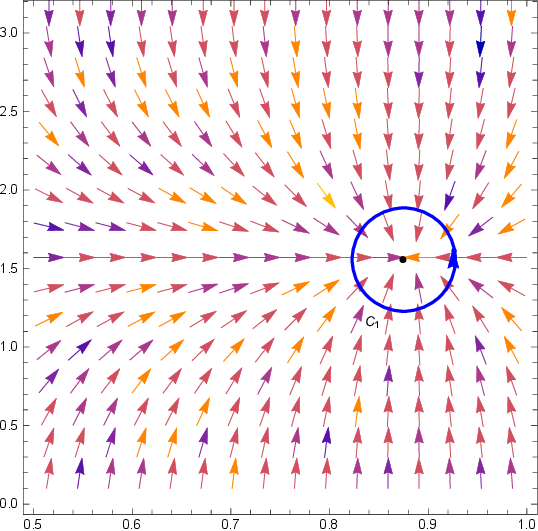}
 \label{6b}}
  \caption{\small{The plot of photon spheres with respect to $  \ell= 1,  a = 0.4,  c = 0.2, \omega = -1$ for $Q>M$ in Fig. (\ref{6a}) and with respect to $\ell = 1,  a = 0.1,  c = 0.7, \omega = -1$ for $Q>M$ in Fig. (\ref{6b})}}
 \label{6}
 \end{center}
 \end{figure}

According to the points mentioned above, we aim to challenge the weak gravity conjecture concerning photon spheres of quantum-corrected (AdS-RN) black holes within Kiselev spacetime, taking into account the Weak Cosmic Censorship Conjecture (WCCC) in relation to the values of various free parameters. In addition to providing comprehensive explanations, we will also succinctly present the results in Table 1. It is known that in the standard case where $ M > Q $, our model should exhibit the typical behavior of a black hole; that is, it possesses a total charge of PS=-1. Building on the aforementioned explanations, we will examine the different modes-$ M > Q $, $ M = Q $, and $ M < Q $-across various parameters. It is crucial to verify the structure in the $ Q > M $ scenario satisfyingly WCCC and photon spheres, ensuring that our structure retains the characteristics of a black hole, namely, it typically has a horizon and a photon sphere with a total negative charge of -1. Therefore, based on the values of the free parameters, we will thoroughly analyze the behavior of different structures as illustrated in the subsequent diagrams. As depicted in Fig. (\ref{2}), when Kiselev spacetime is not considered (i.e., $ c = 0 $), the presence of the quantum correction parameter (a) and the consideration of free parameters such as $ \ell = 1, a = 0.4 $, result in an event horizon for all three cases. The total charge of the photon sphere equals -1, indicating that in the extremal state of a black hole and even when $ Q > M $, our structure upholds the Weak Cosmic Censorship Conjecture (WCCC) and becomes a black hole with a photon sphere (PS) of -1. In this scenario, the photon sphere of the black hole can serve as evidence for the existence of the weak gravity conjecture in the case mentioned. The values considered for the $ Q $ and $ M $ parameters range between 0 and 1 in various cases. In Fig.(\ref{3}), we assume the values $ \ell = 1, a = 0.1, c = 0.2, \omega = 0 $. It is known that for the states $ M > Q $ and $ M = Q $, we observe one and two event horizons, respectively, and the total charge of the photon spheres is -1. However, in the $ Q > M $ case, there is no event horizon, and the total charge of the photon spheres equals 0, signifying a singularity. This is a marked departure from the previous mode. Interestingly, when the value of parameter $ a $ increases from 0.1 to 0.4, our structure exhibits an event horizon, and the total charge of the photon spheres is -1, suggesting that our structure takes the form of a black hole. The photon spheres thus validate the weak gravity conjecture in this instance by satisfying the WCCC, as illustrated in Fig. (\ref{3d}). Furthermore, when we set $ \omega $ to -2.3 in Fig. (\ref{4}), according to the free parameters $ \ell = 1, a = 0.1, c = 0.2 $, our structure assumes a singular form. Under the presumption of an event horizon in $ M > Q $, the total charge of photon spheres is 0, while in the extremal state $ M = Q $ and $ Q > M $, the total charge of the photon spheres is +1, indicating a singularity in these situations
  \begin{figure}[h!]
 \begin{center}
 \subfigure[]{
 \includegraphics[height=4.5cm,width=5cm]{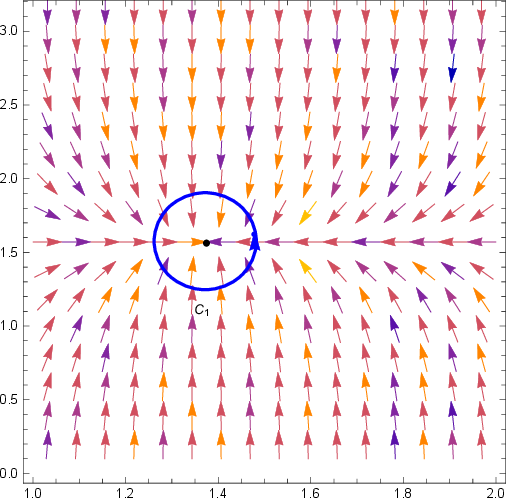}
 \label{5a}}
 \subfigure[]{
 \includegraphics[height=4.5cm,width=5cm]{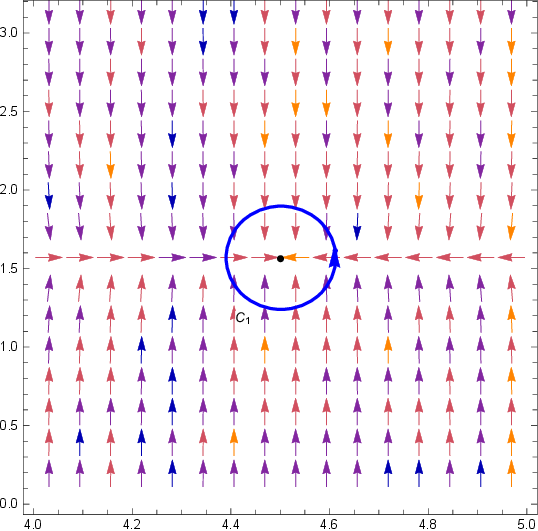}
 \label{5b}}
  \caption{\small{The plot of photon spheres with respect to $ \ell = 1,  a = 0.4,  c = 0.2, \omega = -1/3$ for $Q>M$ in Fig. (\ref{5a}) and with respect to $\ell = 1,  a = 0.1,  c = 0.7, \omega = -1/3$ for $Q>M$ in Fig. (\ref{5b})}}
 \label{5}
 \end{center}
 \end{figure}
We have advanced the conditions of the weak gravity conjecture for the black hole in question, this time considering a different set of parameters, specifically with respect to $\omega = -\frac{1}{3}$. The results are depicted in Fig. (\ref{5}). Generally, for the cases where $M > Q$ and $M = Q$, the behavior is analogous to the previous modes. However, an intriguing observation in this analysis is that for the values of the free parameters ($\ell = 1, a = 0.4, c = 0.2, \omega = -\frac{1}{3}$), in the $Q > M$ scenario, an event horizon is present, and the total topological charge of the photon sphere equals -1. This outcome signifies the structure of a black hole, thereby confirming the validity of the weak gravity conjecture through the photon sphere associated with the specified black hole, as illustrated in Fig. (\ref{4a}). Furthermore, as demonstrated in Fig. (\ref{5b}), reducing the parameter $a$ while increasing $c$ maintains the photon sphere's total topological charge at -1. Consequently, we can preserve the conjecture's validity by decreasing $a$ and augmenting $c$. Also, a similar behavior can be observed in Fig. (\ref{6a}) and Fig. (\ref{6b}), this time for $\omega = -1$ and the mentioned free parameters. It is important to note that for $\omega = -\frac{4}{3}$, the photon sphere generally does not form, or if it does, it is situated within the event horizon. This suggests that such a structure is not applicable for the weak gravity conjecture (WGC).

\section{Conclusions}
In the present study, we have embarked on a journey to study the (WGC) within the realm of quantum-corrected (AdS-RN) black holes situated in Kiselev spacetime. Through the integration of quantum corrections and the unique framework of Kiselev spacetime, we scrutinized the extremality conditions of these black holes. Our results not only fortify the foundation of the WGC but also broaden its scope to encompass the intricate configurations of black holes within Kiselev spacetime. We have shown that, despite the introduction of quantum corrections, the charge-to-mass ratio consistently exceeds the threshold posited by the WGC. This finding is a testament to the conjecture's robustness and its universal applicability. In conclusion, our investigation into the weak gravity conjecture for quantum-corrected black holes within Kiselev spacetime has yielded significant insights. The study confirms that the topological charge of the photon sphere (PS= -1) is a characteristic feature of black holes. The important result is that in $Q>M$, the black hole has photon spheres and satisfies the weak gravity conjecture. Therefore, photon spheres can be a validation for the weak gravity conjecture.
Our analysis across various parameters for different black hole states-$ M > Q $, $ M = Q $, and $ M < Q $-has been crucial in verifying the structure's compliance with the Weak Cosmic Censorship Conjecture (WCCC) and the existence of photon spheres in validation of WGC. Notably, we found that even when $ Q > M $, the structure retains black hole characteristics, such as an event horizon and a photon sphere with a total charge of -1 with respect to upholding the WGC. The study also explores the effects of varying free parameters, including $ \ell, a, c, $ and $ \omega $. Our results indicate that certain parameter configurations, particularly when $ \omega = -\frac{1}{3}, -1$, support the weak gravity conjecture, as evidenced by the presence of an event horizon and a consistent photon sphere. However, for $ \omega = -\frac{4}{3} $, the absence of a photon sphere suggests that the structure does not conform to the conjecture. These findings are critical for understanding the topological aspects of black holes and their photon spheres, providing a new perspective on the weak gravity conjecture within the context of Kiselev spacetime. The detailed results and their implications for the conjecture are presented in Table (\ref{m5}) and illustrated in Fig. (\ref{1}) through Fig. (\ref{5}). Moreover, our exploration illuminates the complex dynamics between quantum mechanics and gravitational forces. The insights gained from this study pave the way for further explorations into the entwined nature of high-energy physics and quantum gravity, potentially unlocking new frontiers in our understanding of the universe. The quantum corrected black hole has been studied in the literature\cite{1,2,3,4,5,6}. However, its phase transition structure may also be scrutinized within various frameworks. One such framework is the study of phase transition structures in topological thermodynamics, a topic that has garnered considerable interest among researchers recently. Another is dynamic phase transition, which offers a broader scope for exploration. There are some attractive points for future research based on extending the analysis of quantum corrections, such as;\\
$\bullet$ Investigating the impact of higher-order quantum corrections on the thermodynamics and phase transition structure of AdS-RN black holes in Kiselev spacetime.\\
$\bullet$ Conduct a comparative study of phase transitions in black holes using topological thermodynamics and traditional approaches.\\
$\bullet$ Analyzing the topological invariants associated with the phase transitions of black holes in Kiselev spacetime.\\
$\bullet$ Examine the dynamic phase transitions of black holes in the context of non-equilibrium thermodynamics.\\
$\bullet$ Studying the time-dependent behavior of phase transitions in black holes and its relation to Hawking radiation.\\
$\bullet$ Analyzing the black holes in modified theories of gravity to evaluate the universality of the WGC and gravitational phenomena and quantum effects.\\
$\bullet$ Exploring the interplay between gravitational anomalies and quantum effects in the vicinity of black holes.\\
$\bullet$ Developing computational models to simulate the quantum-gravitational environment of black holes.\\
$\bullet$ Studying the photon spheres and observational signatures.\\
$\bullet$ Investigate the observational signatures of photon spheres in quantum-corrected black holes and their detectability with current and future telescopes.
\begin{center}
\begin{table}
  \centering
\begin{tabular}{|p{6.2cm}|p{2.5cm}||p{3.5cm}||p{2cm}||}
  \hline
   \hspace{0.8cm}Free parameters  & \hspace{0.5cm} $(Q/M)_{ext}$  & \hspace{0.2cm} PS Total charge & \hspace{0.1cm} PS-WGC \\[3mm]
   \hline
    $ c = 0, \ell = 1,  a = 0.4$ &\hspace{0.5cm} $M>Q$ & \hspace{1.5cm} -1 & \hspace{0.6cm} $\times$ \\[3mm]
   \hline
    $ c = 0, \ell = 1,  a = 0.4$ & \hspace{0.5cm} M = Q & \hspace{1.5cm}  -1 & \hspace{0.6cm} $\times$ \\[3mm]
  \hline
   $ c = 0, \ell = 1,  a = 0.4$ & \hspace{0.5cm} $Q>M$ & \hspace{1.5cm}  -1 & \hspace{0.6cm}$ \surd$\\[3mm]
  \hline
  $\ell = 1,  a = 0.1,  c = 0.2, \omega = 0$ & \hspace{0.5cm} $M>Q$ & \hspace{1.5cm} -1 & \hspace{0.6cm} $\times$\\[3mm]
  \hline
  $\ell = 1,  a = 0.1,  c = 0.2, \omega = 0$ & \hspace{0.5cm} M = Q & \hspace{1.5cm} -1 & \hspace{0.6cm} $\times$ \\[3mm]
  \hline
  $\ell = 1,  a = 0.1,  c = 0.2, \omega = 0$ & \hspace{0.5cm} $Q>M$ &\hspace{1.5cm} 0 & \hspace{0.6cm} $\times$ \\[3mm]
  \hline
  $\ell = 1,  a = 0.4,  c = 0.2, \omega = 0$ & \hspace{0.5cm} $Q>M$ & \hspace{1.5cm} -1 & \hspace{0.6cm} $\surd$ \\[3mm]
  \hline
  $\ell = 1,  a = 0.1,  c = 0.2, \omega = -2/3$ & \hspace{0.5cm} $M>Q$ & \hspace{1.5cm} 0 & \hspace{0.6cm} $\times$ \\[3mm]
  \hline
  $\ell = 1,  a = 0.1,  c = 0.2, \omega = -2/3$ & \hspace{0.5cm} M = Q & \hspace{1.4cm} +1 & \hspace{0.6cm} $\times$ \\[3mm]
  \hline
  $\ell = 1,  a = 0.1,  c = 0.2, \omega = -2/3$ & \hspace{0.5cm} $Q>M$ & \hspace{1.4cm} +1 & \hspace{0.6cm} $\times$ \\[3mm]
  \hline
  $\ell = 1,  a = 0.4,  c = 0.2, \omega = -1/3$ & \hspace{0.5cm} $Q>M$ & \hspace{1.5cm} -1 & \hspace{0.6cm} $\surd$ \\[3mm]
  \hline
  $\ell = 1,  a = 0.1,  c = 0.7, \omega = -1/3$ & \hspace{0.5cm} $Q>M$ & \hspace{1.5cm} -1 & \hspace{0.6cm} $\surd$ \\[3mm]
  \hline
  $\ell = 1,  a = 0.4,  c = 0.2, \omega = -1$ & \hspace{0.5cm} $Q>M$ & \hspace{1.5cm} -1 & \hspace{0.6cm} $\surd$ \\[3mm]
  \hline
  $\ell = 1,  a = 0.1,  c = 0.7, \omega = -1$ & \hspace{0.5cm} $Q>M$ & \hspace{1.5cm} -1 & \hspace{0.6cm} $\surd$ \\[3mm]
  \hline
\end{tabular}
\caption{Summary of the results}\label{m5}
\end{table}
 \end{center}

\end{document}